\documentclass{moriond}

\usepackage{graphicx}
\usepackage{xspace}
\usepackage{multirow}
\usepackage{comment}
\usepackage{amsmath}
\usepackage{braket}

\def\be{\begin{equation}}
\def\ee{\end{equation}}
\def\bea{\begin{eqnarray}}
\def\eea{\end{eqnarray}}

\makeatletter
\newcommand{\oset}[3][0ex]{%
  \mathrel{\mathop{#3}\limits^{
      \vbox to#1{\kern-2.5\ex@
        \hbox{$\scriptstyle#2$}\vss}}}}
\makeatother

\newcommand\overbar[1]{\oset[-0.2ex]{%
    \textbf{--}}{#1}}

\newcommand{\nue}{\ensuremath{\nu_{e}}\xspace}
\newcommand{\nuecc}{\ensuremath{\nu_{e}}~CC\xspace}
\newcommand{\nuebar}{\ensuremath{\overbar{\nu}_e}\xspace}
\newcommand{\nuebarcc}{\ensuremath{\overbar{\nu}_e}~\text{CC}\xspace}

\newcommand{\numu}{\ensuremath{\nu_{\mu}}\xspace}
\newcommand{\numucc}{\ensuremath{\nu_{\mu}}~\text{CC}\xspace}
\newcommand{\numubar}{\ensuremath{\overbar{\nu}_\mu}\xspace}
\newcommand{\numubarcc}{\ensuremath{\overbar{\nu}_\mu}~\text{CC}\xspace}

\newcommand{\dmsq}{\ensuremath{\Delta m^2_{32}}\xspace}
\newcommand{\snsq}{\ensuremath{\sin^2 \theta_{23}}\xspace}
\newcommand{\deltacp}{\ensuremath{\delta_{\rm{CP}}}\xspace}

\newcommand{\Photo}{}

\begin{document}
\vspace*{4cm}
\title{RECENT RESULTS FROM NOvA}

\author{ E. CATANO-MUR (for the NOvA Collaboration)}

\address{Department of Physics, William \& Mary,\\
  Williamsburg, Virginia 23187, USA}

\maketitle\abstracts{
We present the most recent 3-flavor neutrino oscillation results from the NOvA long-baseline experiment, using a joint fit of $\numu \to \numu$, $\numubar \to \numubar$, $\numu \to \nue$, and $\numubar\to \nuebar$ channels, and an accumulated exposure of $13.6 \times 10^{20}$ protons-on-target of neutrino beam and $12.5 \times 10^{20}$ protons-on-target of antineutrino beam.
The best-fit values for the atmospheric parameters are $\dmsq = (2.41\pm0.07)\times 10^{-3}$~eV$^2$, $\snsq = 0.57^{+0.03}_{-0.04}$ and $\deltacp = (0.82^{+0.27}_{-0.87}) \pi$. The data disfavor combinations of oscillation parameters that lead to large asymmetries between the rates of \nue vs \nuebar appearance. 
}
\section{Introduction}

Neutrino oscillations are transitions in-flight between the flavor neutrinos, caused by non-zero neutrino masses and neutrino mixing. 
In the 3-flavor paradigm, the flavor eigenstates ($\nue,\numu,\nu_\tau$) are linear combinations of three mass eigenstates ($\nu_1, \nu_2, \nu_3$), 
\begin{align}
     | \nu_{\alpha} \rangle  = \sum_{i=1}^{3} U^*_{\alpha i} | \nu_i \rangle,
     \qquad \qquad \alpha = e, \mu, \tau,
\end{align}
where $U$ is the unitary mixing matrix. 
The probabilities of flavor transitions will depend on: the elements of the mixing matrix, parameterized by three angles ($\theta_{12}, \theta_{13}, \theta_{23}$) and one phase ($\delta_{CP}$); the squared differences between mass eigenvalues ($\Delta m^2_{21},\Delta m^2_{32}$); the energy of the neutrino ($E$), and the distance it traveled ($L$). 

Evidence of neutrino oscillations has been onserved in solar, atmospheric, reactor and accelerator experiments. The different types of neutrino sources and experimental setups have characteristic values of $L$ and $E_\nu$, which set the ranges of $|\Delta m^2|$ and the kind of oscillations to which they are sensitive~\cite{Zyla:2020zbs}.
In particular, long-baseline accelerator neutrino experiments have $L/E\sim \text{500 GeV/km}$, and are sensitive to the atmospheric-sector oscillations with a $\Delta m^2 \sim {2.5\times 10^{-3}}{\text{ eV}^2}$.
With a primordially muon neutrino (or antineutrino) beam, the combined measurements of long-baseline disappearance ($\numu \to \numu$ and $\numubar \to \numubar$) and appearance ($\numu \to \nue$ and $\numubar \to \nuebar$) allow the estimation of $ \theta_{23}$ and $\Delta m^2_{32}$, and the determination of the neutrino mass ordering ($\text{sign}(\dmsq)$), the octant of $\theta_{23}$, and charge-parity (CP) violation in the neutrino sector.

\section{The NOvA experiment}
\begin{figure}
    \centerline{\includegraphics[width=0.85\textwidth]{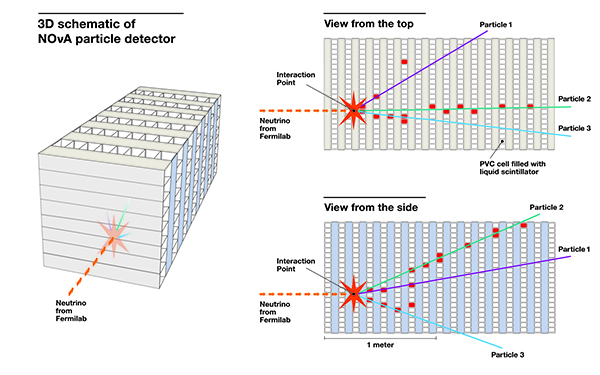}}
    \caption[]{Schematic of the NOvA detectors showing the alternating
    plane orientations. The combination of top and side views 
    offers 3D reconstruction of the
    particle trajectory~\cite{ref:vms}. }
    \label{fig:3d}
\end{figure}

NOvA 
consists of two finely-segmented liquid-scintillator detectors operating 14.6 mrad off-axis from Fermilab’s NuMI muon neutrino (or antineutrino) beam~\cite{Adamson:2015dkw}. The detectors comprise PVC cells filled with a liquid scintillator, arranged in planes that alternate between horizontal and vertical directions to allow 3D reconstruction, as shown in Fig. \ref{fig:3d}.
 The near detector (ND) has 214 of such planes, and  samples the
beam  1 km from the source. The far detector (FD) has 896 planes, and observes
the oscillated beam 810 km downstream, near Ash River, MN.

\begin{figure}
    \centerline{\includegraphics[width=0.8\textwidth]{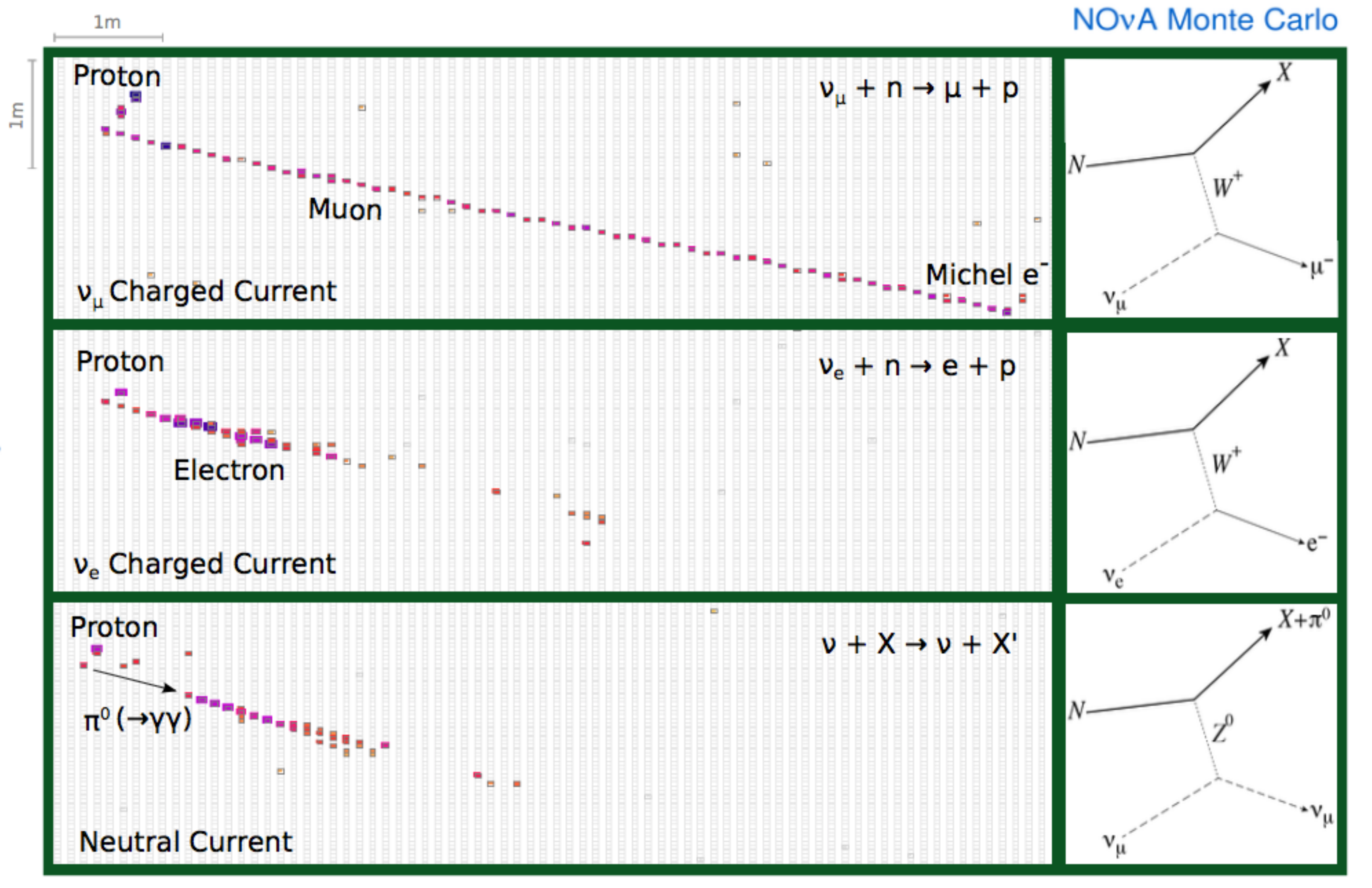}}
    \caption[]{Simulated neutrino interactions in the NOvA detectors: \numucc
               (top), \nuecc (middle), and NC (bottom)~\cite{Baird:2015pgm}.}
    \label{fig:topol}
\end{figure}

The data is collected in the detectors in the form of individual hits, which get clustered based on proximity in time and space to construct event candidates.
These clusters are then processed to estimate the location of the interaction vertex, its direction, and other reconstructed quantities. Further, a convolutional neural network is used to classify the neutrino event candidates into \nuecc, \numucc, NC, or cosmogenic backgrounds. Figure \ref{fig:topol} shows  characteristic topologies of CC and NC events in the detectors. 

The NOvA detectors are tracking calorimeters, and both characteristics are used to estimate the energy of the neutrino candidates. For \numucc events, the energy of the muon is estimated using the track length ($\sim$4\% resolution), and the energy of the hadronic system is estimated using calorimetry ($\sim$30\% resolution). The \nuecc energy is a function of the electromagnetic energy ($\sim$10\% resolution)   
and the hadronic energy, where both use calorimetry.

\section{The NOvA oscillation analysis}

Four channels are relevant to the estimation of neutrino oscillation parameters with NOvA: $\numu \to \numu$, $\numubar \to \numubar$, $\numu \to \nue$, and $\numubar\to \nuebar$. The predictions of the spectra at the FD are constructed using a data-driven approach, using high-statistics measurements with the ND to improve the base simulation and constrain systematic uncertainties. For both appearance and disappearance, the signal predictions are corrected using \numucc ND samples, for neutrino or antineutrino beam mode separately. The \nuecc ND samples are used to correct the background predictions for the appearance channel. 
 
\begin{figure}
\centerline{
  \includegraphics[width=0.42\linewidth]{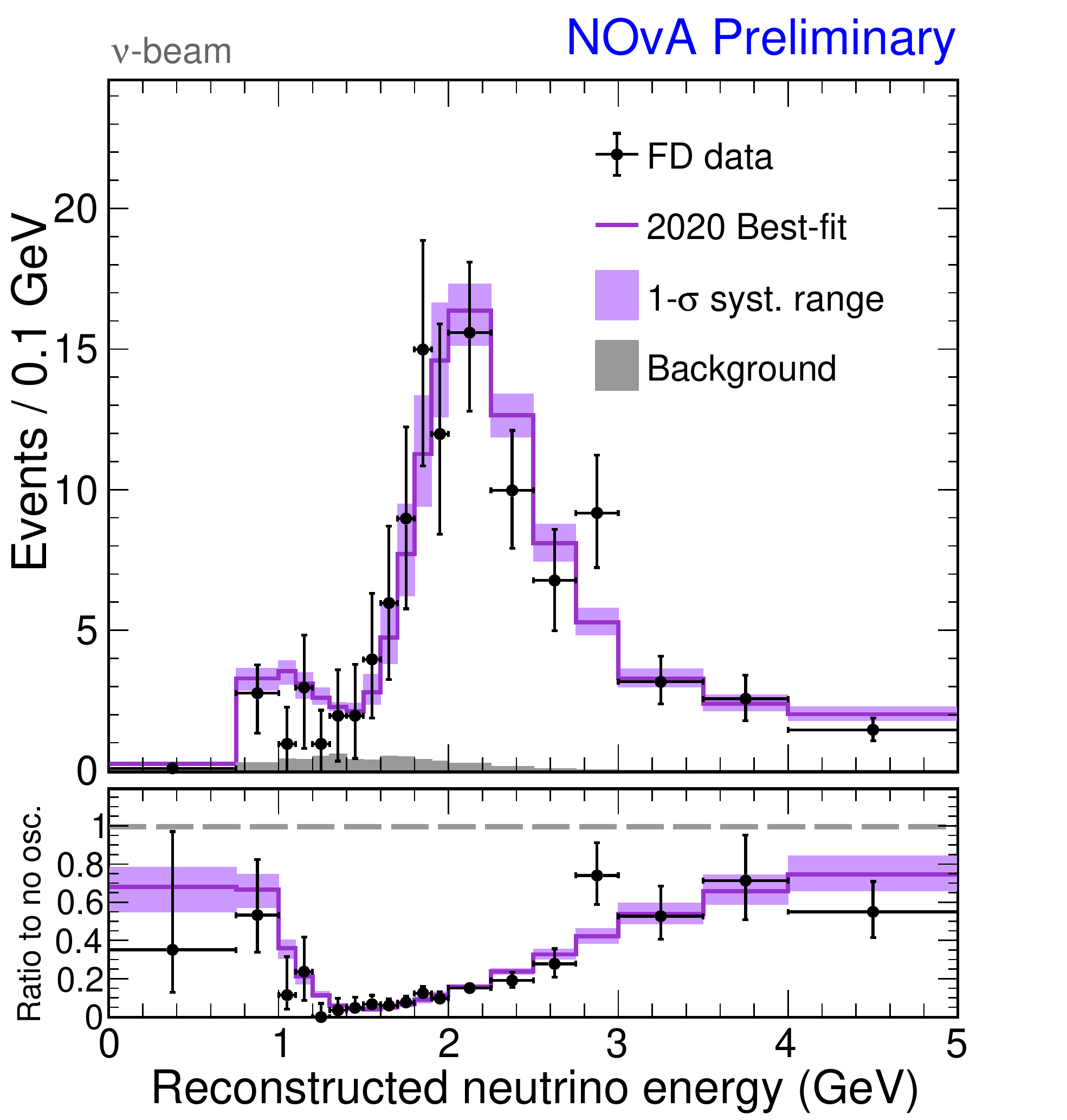}
  \includegraphics[width=0.42\linewidth]{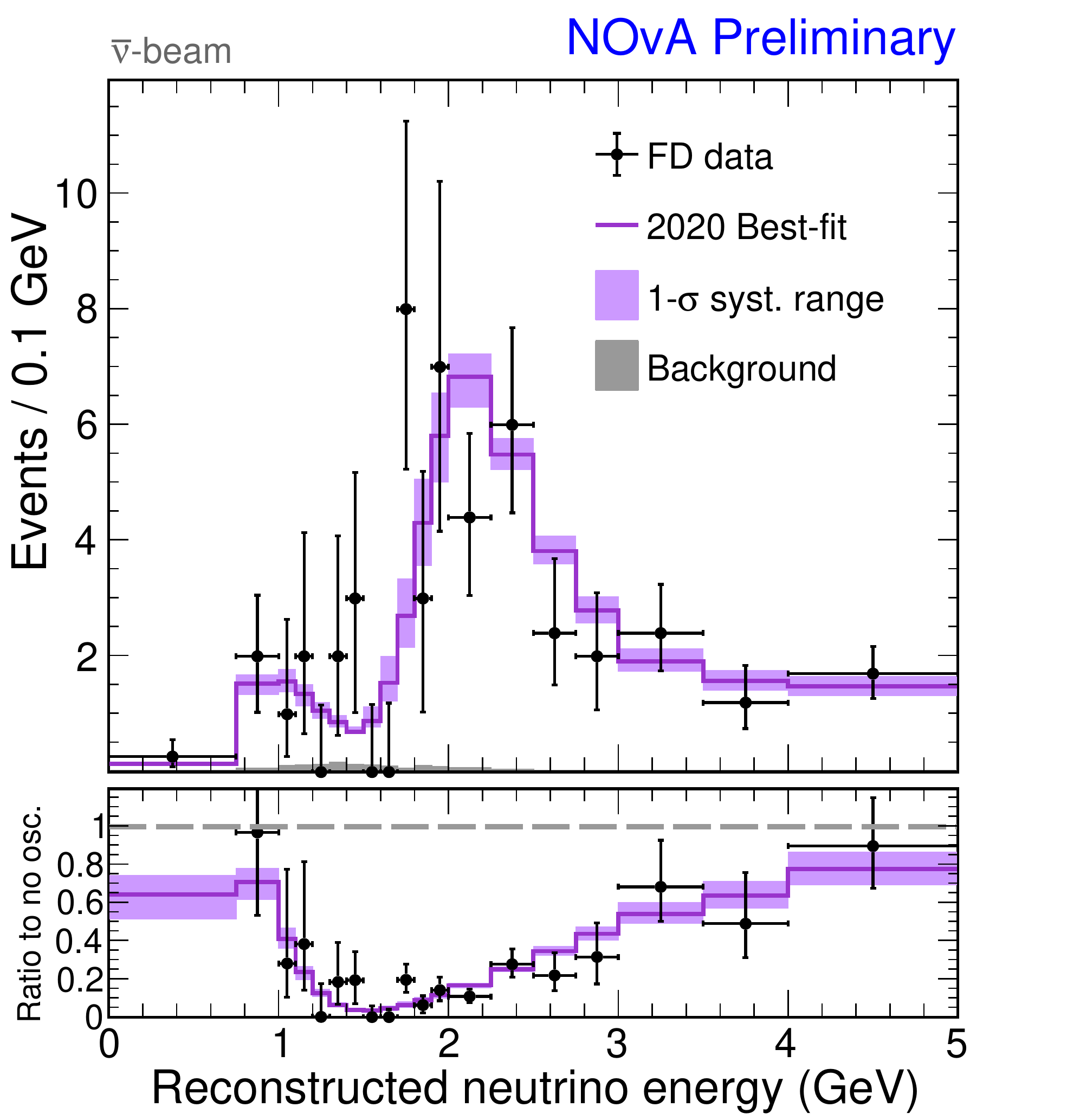}
  }
  \caption[]{Reconstructed neutrino energy spectra for the \numucc (left) and \numubarcc (right) samples at the FD.}
  \label{fig:numu_fd}
\end{figure}

\begin{figure}
\centerline{
  \includegraphics[width=0.5\linewidth]{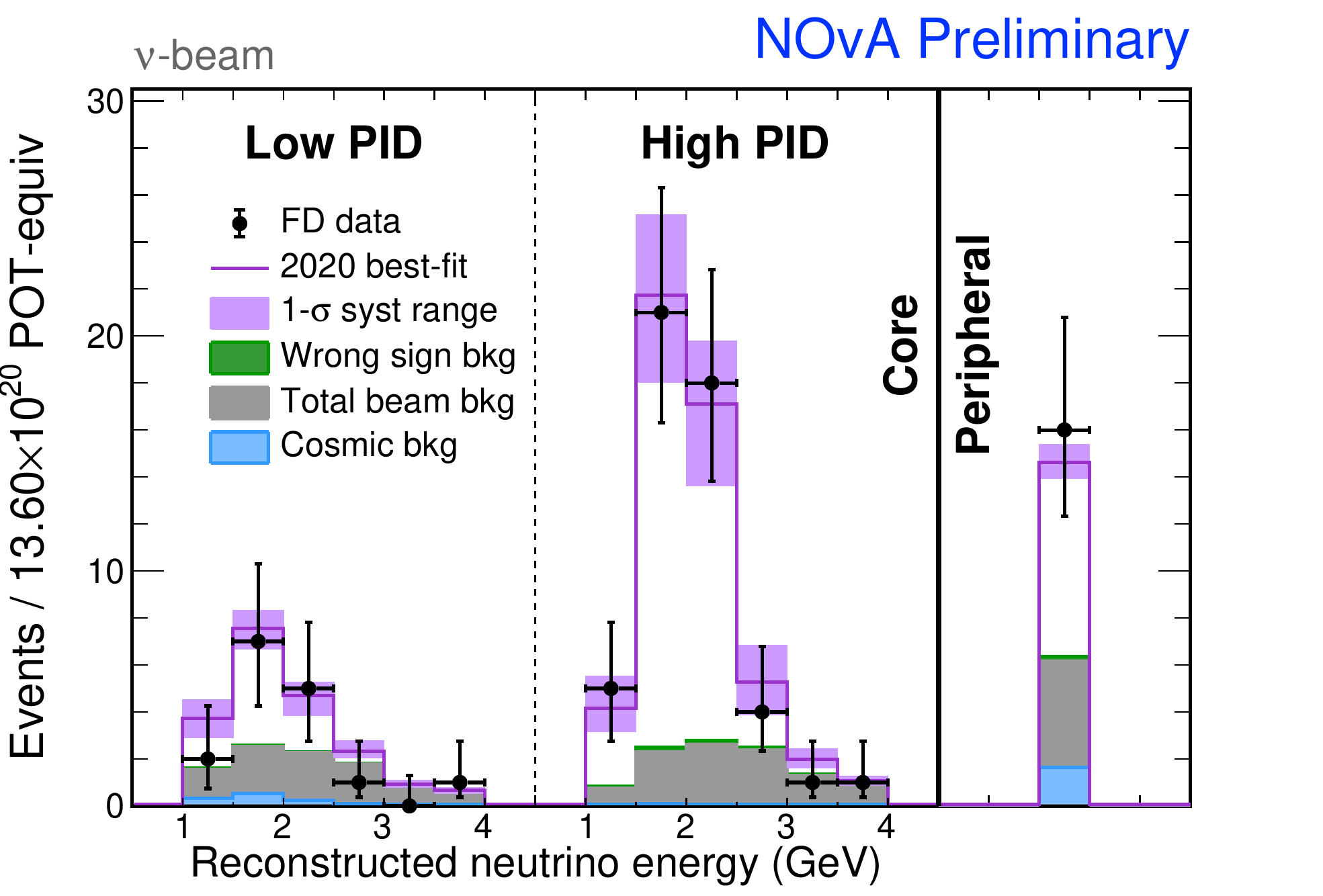}
  \includegraphics[width=0.5\linewidth]{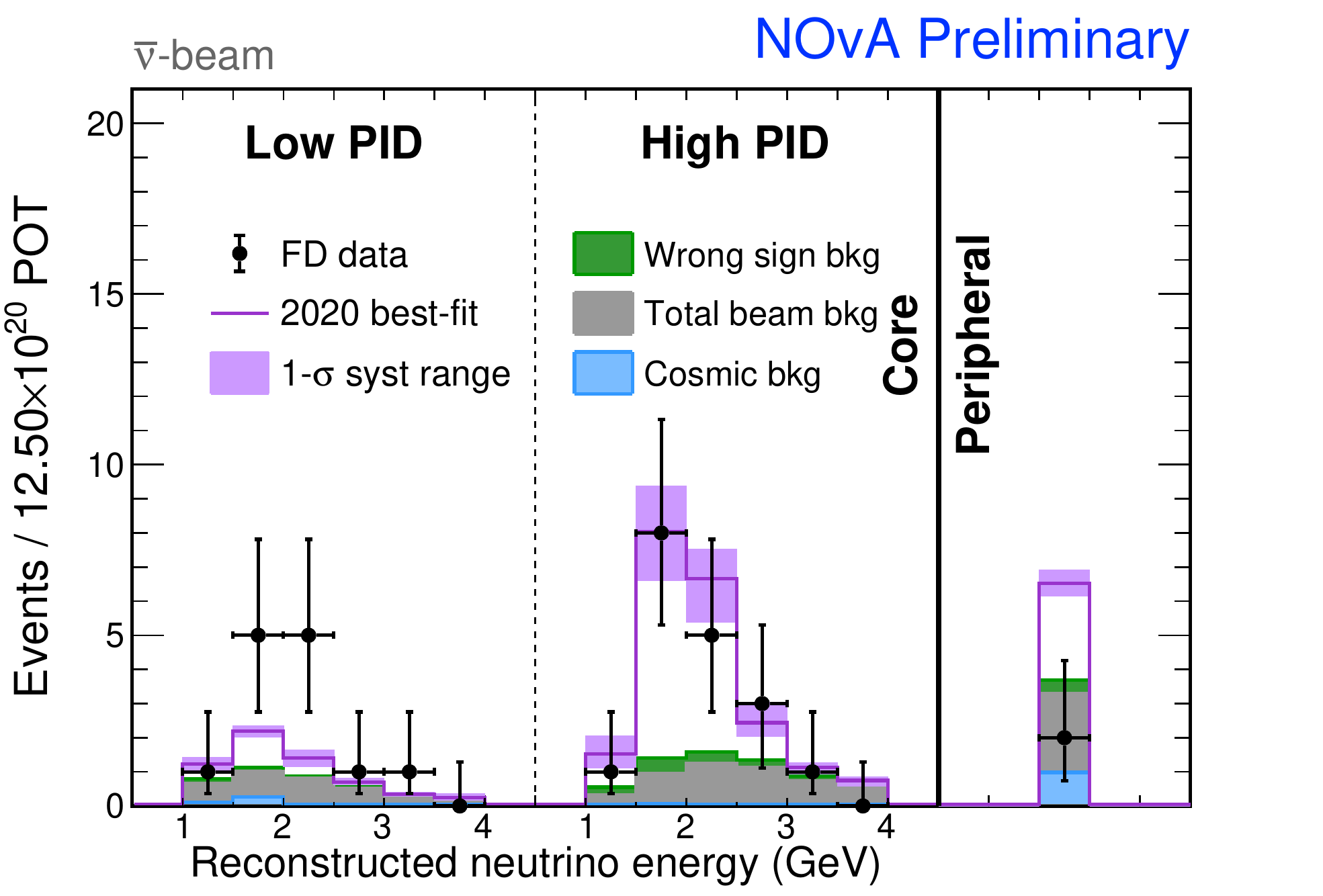}
  }
  \caption[]{Reconstructed neutrino energy spectra for the \nuecc (left) and \nuebarcc (right) samples at the FD.}
  \label{fig:nue_fd}
\end{figure}

NOvA’s latest measurements of neutrino oscillation parameters~\cite{NOvA:2021nfi} use data recorded between 2014 and 2020 and correspond to $13.6 \times 10^{20}$ protons-on-target of neutrino beam and $12.5 \times 10^{20}$ protons-on-target of antineutrino beam. Figures \ref{fig:numu_fd} and \ref{fig:nue_fd} show the energy spectra of the \numucc, \numubarcc, \nuecc and \nuebarcc candidates observed in the FD, compared to the best fit predictions. Table \ref{tab:event-count} summarizes the event counts for each sample, comparing the observations with the estimated signal, background and total predictions. 

\begin{table}[h]
  \caption{
    Event counts at the FD, both observed and predicted at the best-fit point (Eq. \ref{eq:bestfit}).}
  \label{tab:event-count} \centering \renewcommand{\arraystretch}{1.2} \setlength{\tabcolsep}{6pt}
  \begin{tabular}{|l|cc|cc|}
    \hline
    & \multicolumn{2}{c|}{Neutrino beam} & \multicolumn{2}{c|}{Antineutrino beam} \\
    & {\numucc}             & {\nuecc}              & {\numubarcc}          & {\nuebarcc} \\
    \hline
    Signal                          & 214.1$^{+14.4}_{-14.0}$ & 59.0$^{+2.5}_{-2.5}$  & 103.4$^{+7.1}_{-7.0}$ & 19.2$^{+0.6}_{-0.7}$ \\
    Background                      &   8.2$^{+1.9}_{-1.7}$ & 26.8$^{+1.6}_{-1.7}$  &   2.1$^{+0.7}_{-0.7}$ & 14.0$^{+0.9}_{-1.0}$  \\
    \hline
    Best fit                        & 222.3                 & 85.8                  & 105.4                 & 33.2 \\
    Observed                        & 211                   & 82                    & 105                   & 33   \\
    \hline
  \end{tabular}
\end{table}

The best-fit parameters are obtained by minimizing a Poisson negative log-likelihood ratio between oscillated FD predictions and the observed spectra~\cite{NOvA:2021nfi}.
The fit includes three parameters that are allowed to vary freely ($ \Delta m^2_{32}, \sin^2 \theta_{23},\delta_{CP} $); 67 systematic uncertainties; and constraints from solar and reactor experiments  ($\Delta m^2_{21} = 7.53  \times 10^{-5}~{\rm eV}^2$, $\sin^2\theta_{12} =0.307 $, and $\sin^2\theta_{13} = 0.0210 \pm 0.0011$). The resulting best-fit oscillation parameters (and their $1\sigma$ allowed ranges) are:  \begin{align} \label{eq:bestfit}
    \Delta m^2_{32} &= (+2.41 \pm 0.07) \times 10^{-3} \text{eV}^2, \\
    \nonumber\sin^2 \theta_{23} & = 0.57^{+0.03}_{-0.04}, \\
    \nonumber\delta_{CP} &= (0.82^{+0.27}_{-0.87}) \pi.
\end{align}

The data disfavor combinations of parameters that lead to a strong asymmetry in the rate of \nue vs \nuebar appearance: $\deltacp = \pi/2$  is excluded at more than 3\,$\sigma$ in the inverted mass ordering, and $\deltacp = 3\pi/2$ in the normal ordering  is disfavored at 2\,$\sigma$ confidence. Considering all permutations of the mass ordering and the octant, degeneracies are such that all values of \deltacp are compatible with the data, with no preference of CP conservation over CP violation~\cite{NOvA:2021nfi}.

\begin{figure}
\centerline{
  \includegraphics[width=0.5\linewidth]{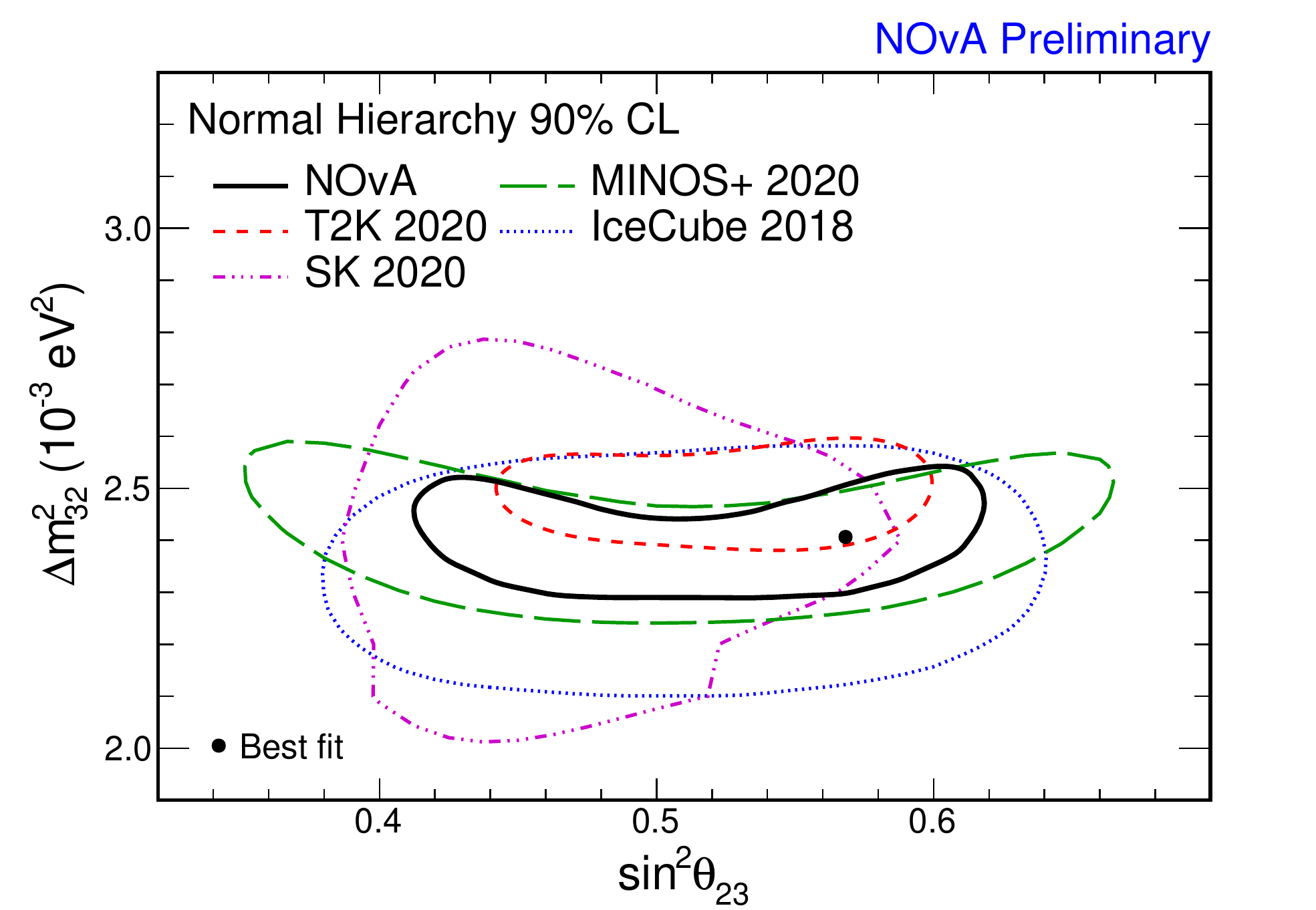}
  \includegraphics[width=0.5\linewidth]{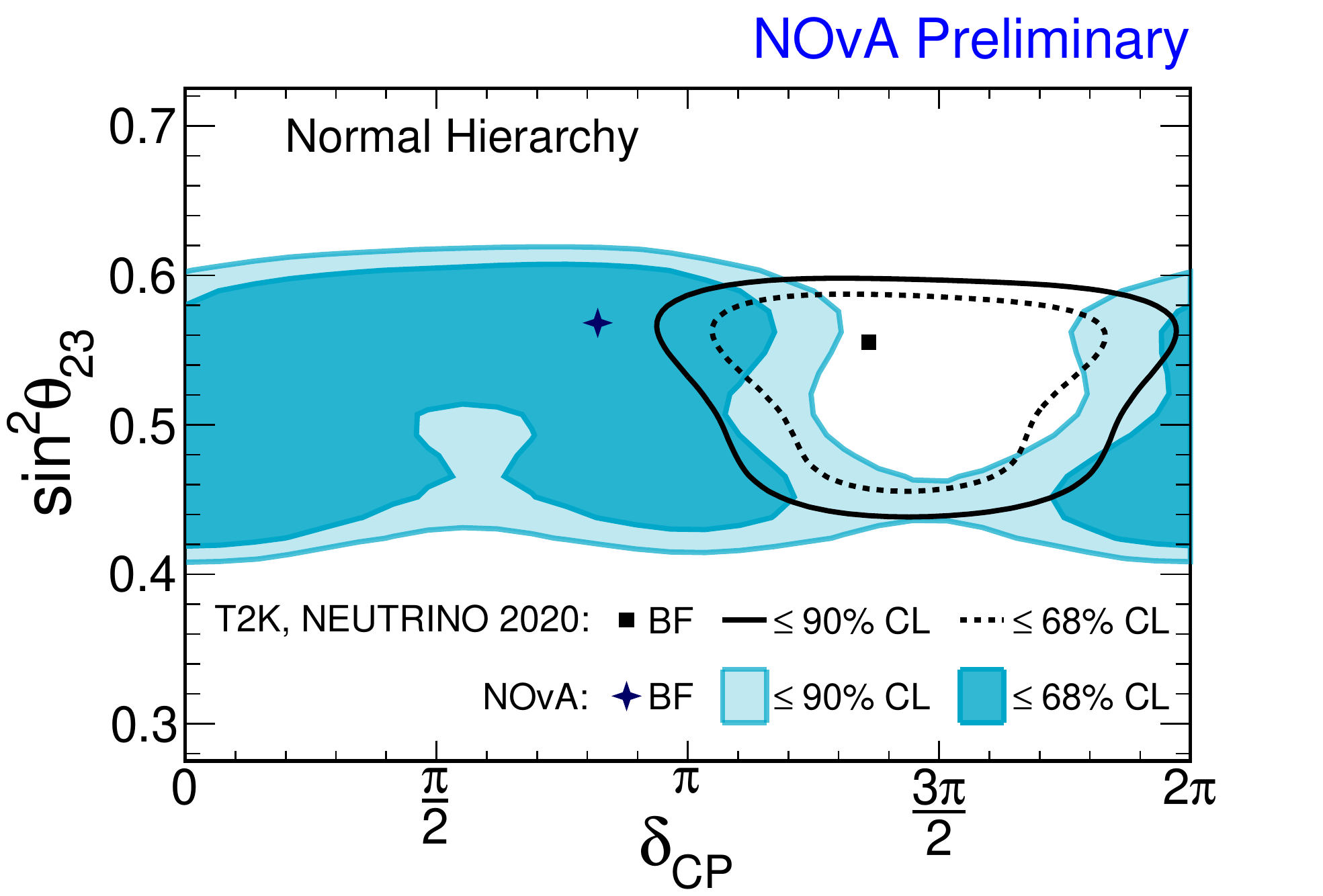}
  }
  \caption[]{Left: The 90\% confidence level region for \dmsq and \snsq,
   comparing the NOvA allowed region (black) with the contours from other experiments~\cite{T2K:2021xwb,Super-Kamiokande:2020,MINOS:2020llm,IceCube:2017lak}.
   Right: The 68\% and 90\% confidence level regions for \snsq vs. \deltacp in the normal mass ordering, comparing the NOvA allowed regions (color areas) with the T2K contours~\cite{T2K:2021xwb}.
   }
   \label{fig:contours}
  \end{figure}

Figure \ref{fig:contours} shows 2-dimensional allowed regions for NOvA compared to other experiments, for the normal mass ordering. The $\dmsq \times \snsq$ contours show consistency among atmospheric and accelerator neutrino oscillation experiments. In the $\snsq \times \deltacp$ contours, the best-fit point of the T2K experiment lies in a region disfavored by this analysis, but regions of overlap remain~\cite{NOvA:2021nfi}. This purported tension is not observed in the inverted mass ordering. The T2K and NOvA collaborations are working towards a joint neutrino oscillation analysis using both experiments’ data sets simultaneously~\cite{ref:nova-t2k}.

\begin{figure} 
    \centerline{
        \includegraphics[width=0.48\textwidth]{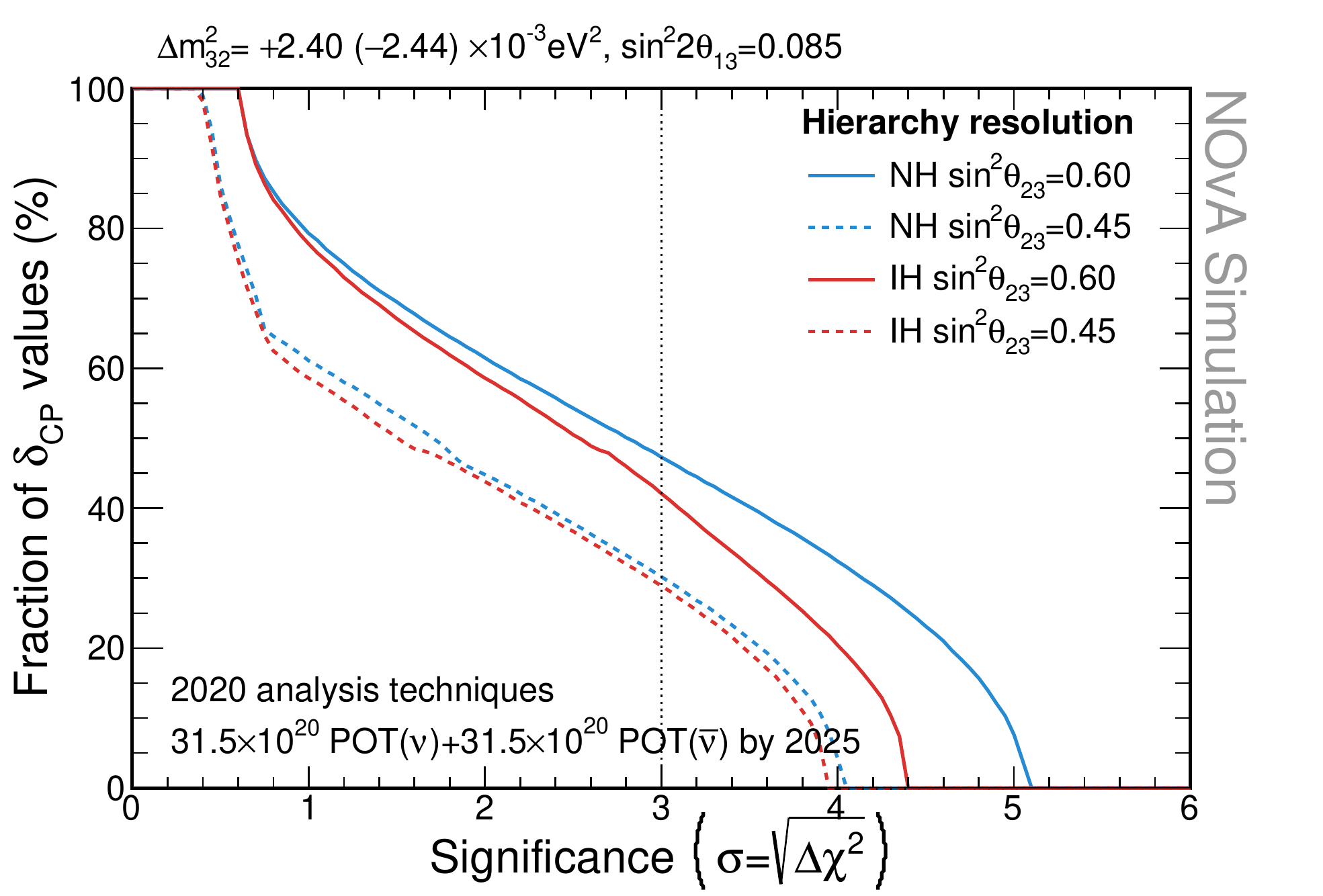}
        \includegraphics[width=0.48\textwidth]{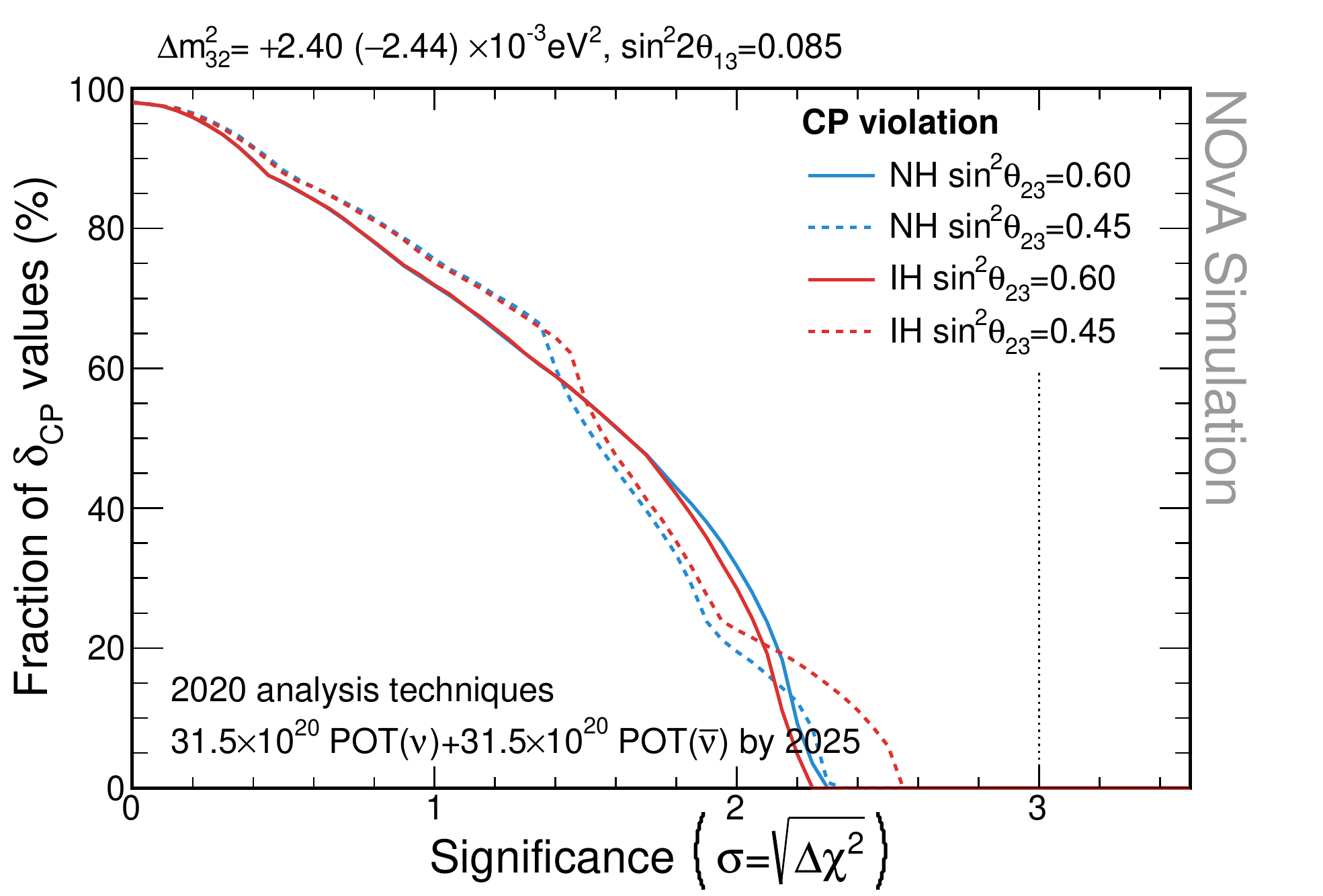}
    }
    \caption[]{
    Fraction of \deltacp values in the range $[0,2\pi)$  for which the neutrino mass ordering (left) or CP non-conservation (right) could be resolved by NOvA at a given sensitivity $\sigma$ by the year 2025, assuming an accumulated exposure of $63\times 10^{20}$ protons-on-target.
    The lines correspond to different true values of $\sin^2 \theta_{23}$ and true normal ordering (blue) or inverted ordering (red).
    }
    \label{fig:reach}
\end{figure}

NOvA is expected to take data through 2026, and the current projection for the ultimate exposure is 60-70$\times 10^{20}$ protons-on-target, approximately doubling the data analyzed so far~\cite{Shanahan:2021jlp}. Figure \ref{fig:reach} shows some projections for the resolution of the neutrino mass ordering and CP violation.
NOvA could reach $2\sigma$ determination of CP violation, $3\sigma$ sensitivity to the ordering for 30-50\% of \deltacp values, and up to $\sim 5 \sigma$ in the most favorable case. 

\section*{Acknowledgments}

This document was prepared by the NOvA collaboration using the resources of the Fermi National Accelerator Laboratory (Fermilab),
a U.S. Department of Energy, Office of Science, HEP User Facility. Fermilab is managed by Fermi Research Alliance, LLC (FRA),
acting under Contract No. DE-AC02-07CH11359. This work was supported by the U.S. Department of Energy; the U.S. National Science Foundation; the Department of Science and Technology, India; the European Research Council; the MSMT
CR, GA UK, Czech Republic; the RAS, RFBR, RMES, RSF, and BASIS Foundation, Russia;
CNPq and FAPEG, Brazil; STFC, UKRI, and the Royal Society, United Kingdom; and
the State and University of Minnesota.
We are grateful for the contributions of the staffs of the University of Minnesota at the Ash River Laboratory and of Fermilab.

\end{document}